\documentclass[twoside]{article}

\usepackage{titling} 
\usepackage{hyperref} 
\usepackage{authblk}
\usepackage[T1]{fontenc}
\usepackage[utf8]{inputenc}
\usepackage{slashed}
\usepackage{dsfont}
\usepackage{amsfonts}
\usepackage{amsmath}
\usepackage{cleveref}

\usepackage{blindtext} 
\usepackage[dvipsnames]{xcolor}
\usepackage[sc]{mathpazo} 
\usepackage[T1]{fontenc} 
\usepackage{microtype} 

\usepackage[english]{babel} 

\usepackage[hmarginratio=1:1,top=32mm,columnsep=20pt]{geometry} 
\usepackage[hang, small,labelfont=bf,up,textfont=it,up]{caption} 
\usepackage{booktabs} 

\usepackage{lettrine} 

\usepackage{enumitem} 
\setlist[itemize]{noitemsep} 

\usepackage{abstract} 

\usepackage{titlesec} 
\renewcommand\thesection{\Roman{section}} 
\renewcommand\thesubsection{\roman{subsection}} 
\titleformat{\section}[block]{\large\scshape\centering}{\thesection.}{1em}{} 
\titleformat{\subsection}[block]{\large}{\thesubsection.}{1em}{} 

\usepackage{fancyhdr} 
\usepackage{cleveref}

\begin{document}

\begin{center}

{\Large\bf Planar generalized electrodynamics for one-loop amplitude in the Heisenberg picture}
\end{center}

\bigskip

\begin{center}

{\large D. Montenegro$^{a}$ and
 B. M. Pimentel$^{a}$
} 

\vskip 0.3cm

{\em $^a$S\~{a}o Paulo State University (UNESP), Institute for Theoretical Physics (IFT), \\ R. Dr. Bento Teobaldo Ferraz 271, BR} \\

\end{center}

\bigskip

\smallskip

\noindent

\begin{abstract}
We examine the generalized quantum electrodynamics as a natural extension of the Maxwell electrodynamics to cure the one-loop divergence. We establish a precise scenario to discuss the underlying features between photon and fermion where the perturbative Maxwell electrodynamics fails. Our quantum model combines stability, unitarity, and gauge invariance as the central properties. To interpret the quantum fluctuations without suffering from the physical conflicts proved by Haag's theorem, we construct the covariant quantization in the Heisenberg picture instead of the Interaction one. Furthermore, we discuss the absence of anomalous magnetic moment and mass-shell singularity.
\end{abstract}


\hfill

\hfill   
 
\section{Introduction} 
 
Quantum electrodynamics $(QED_4)$ is a success theory from the theoretical and experimental aspects, which rules out the quantum properties of spinor-photon interaction. As we known, dimensionality is a feature which arises new phenomena in the $QED$. In $(2+1)$ dimensions, we realize the photons are free and strongly coupled in the ultraviolet and infrared regions of the spectrum, respectively. Furthermore, a detailed investigation suggests $QED_3$ as a relevant model to interpret $QCD_4$ \cite{QCD}, macroscopic quantum Hall effect \cite{Hall},  ultra-cold atoms \cite{opletter}, and theoretical applicability on high$-T_c$ cuprate superconductors \cite{su3perco}. It can also be used to study chiral symmetry by Schwinger-Dyson equations \cite{Lo2010fm}.

It is well-known the adding higher derivative corrections to the original theory is a process for constructing a more fundamental one \cite{referee2}. In recent years, effective field theory models have shown progress in many physical contexts since they can incorporate different energy scales. They are strongly recommended due to draw new perspectives over underlying physical aspects. It is instructive to note the lower and higher derivative theories have distinct nature. The latter increases the family of solutions, which means the former was less accurate. To be more precise, the solution domains show a fundamental characteristic of the system. We shall stress Ostrogradsky's theorem already proved non-degenerate higher derivative theories present at least a linear instability \cite{Ostrogradski}. This argument, which prevents a canonical quantization, is responsible for negative energy modes and vacuum instability. The first question we need to ask is how we demand a sensible procedure to avoid these inconsistencies. Up to now, much progress has been made to clarify in what sense the positivity notion of the canonical Noether's energy is indispensable to determine stability. What has not been explained is that this definition left out several higher order theories. To elucidate what means the unbounded energy spectrum from Noether's theorem, we shall look more closely at the canonical energy. To understand this difference, Kaparulin \textit{et al.} offer an alternative to enlarge the stability concept where the Lagrangian anchor associates the bounded integral of motion with the time-translational invariance \cite{russo}.  

   
The benefit of higher derivative theory reveals further advances in our understanding of the Maxwell electromagnetism. Podolsky incorporated a second order gauge derivative in the $QED_4$ \cite{Podolskyorigin} to solve the old classical issues as the $r^{-1}$ singularity \cite{Podolskyorigin} and famous $"4/3"$ problem  \cite{43problem}. This theory sheds new light not only on solving the problems already mentioned but also suppressing quantum divergences. Consequently, it also motivates to investigate the new perspectives of the so-called generalized quantum electrodynamics $(GQED_3)$. Far from being redundant, the $GQED_3$ is the unique second order gauge theory that preserves the linearity, Lorentz, and gauge invariance, apart of a total derivative term \cite{Cuzinatto}. Moreover, we are able to construct a covariant quantization to the $GQED_3$ without inconsistency since the stability \cite{russo} and unitary (BRSS symmetry) \cite{vsfd} were demonstrated.

The standard perturbative formalism to quantum field theory $(QFT)$ has archived great success in extracting physical predictions from the cross section experiments. According to the conventional physical interpretation, the Interaction picture is the basis for exploring non-trivial quantum phenomena. The question surrounding this picture is that we cannot describe an interacting system \cite{Haag}. In essence, if the Interaction picture assumes a Fock representation for asymptotic times, governed by the free Hamiltonian $H_0$, the interacting Hamiltonian $H_{int}$ cannot annihilate the free vacuum, where elementary phenomena occur as vacuum polarization. In other words, for a finite time, there is no physical state in Interaction picture, which relates by a unitary equivalent transformation to a Fock space. We stress these pathological issues are not a mathematical sophistication but concerned with the $QFT$ foundations \cite{Haag2}.

For the reasons already presented, we approach the subject of the $GQED_3$ to calculate the radiative corrections in the Heisenberg picture. We apply Kállën's framework as an alternative formulation to describe the interacting fields \cite{Kallen1,Kallen2}. The recipe for this perturbative analysis is an adequate strategy to explore the content of the $QFT$. That is to say, the principal quantum objects are not Green's functions but rather dynamical equations, which enables us to work with the total Hamiltonian. In this paper, we will show the Kállën's method attracts attention because of the simplicity. Other applications of the Heisenberg picture are found in the Thirring Model \cite{Lunardi}, vacuum polarization \cite{Tomazelli4}, and non-perturbative mass generation \cite{rabanal} in the $QED_3$.

The paper is organized as follows. In section \ref{deux}, we proceed with a brief introduction of Podolsky's structure with the relevant propagators of the $GQED_3$. In section \ref{444}, beyond defining the perturbative approach in the Heisenberg picture, we present the aspects to understand how the radiative corrections are consistent with the $QFT$ principles. We show the vertex and electron self-energy corrections exhibit a finite loop in section \ref{trois} and \ref{quatre}, respectively. Finally, we present our discussion and final remarks in section \ref{doendo}. At the end of the paper, the appendix \ref{appa} points out technical details.

\hfill

\section{Formalism}\label{deux}

In this section, we review the basic structure of $GQED_3$ to interpret the significance of the Podolsky contribution. By considering the lagrangian written as\footnote{We adopt $x_3 = i x_o = i c t $ and $ - ds^2 = dx^2_\alpha $. The Greek indices run from $1$ to $3$ with natural units $\hbar = c = 1$.}

\begin{equation}\label{llkk}
\mathcal{L}_{GQED_3} = -\frac{1}{4} F^{\mu\nu} F_{\mu\nu} - \frac{a^2}{2} \partial^\mu F_{\mu\beta} \partial_\alpha F^{\alpha\beta},  
\end{equation}
where the strength field is $F_{\mu\nu} = \partial_\mu A_\nu - \partial_\nu A_\mu$ with $A_\mu$ and its derivative continuous up to fourth order. The $e$ and $m_p = a^{-1} $ are the gauge coupling and Podolsky parameter\footnote{The $m_P$ is not a gauge parameter but rather a physical quantity. We can measure by suitable experiments. In $(3+1)$ spacetime dimensions, the experiments to detect proposed are \cite{Cuzinatto2011zz}.}, respectively. The equations of motion are

\begin{equation}\label{elpodd}
(1 - a^2 \Box ) \partial_\mu F^{\mu \nu} = 0.
\end{equation}
As a preliminary to specify a physical gauge theory, we shall impose restrictions on the dynamical gauge fields to reduce the redundant variables. The first convenient guess was the known Lorentz gauge $\Omega[A] = \partial_\mu A^\mu$, proposed by Podolsky \cite{Podolskyorigin}. Even though it would imply a reasonable choice, this restriction cannot fix the gauge since the longitudinal components of \eqref{elpodd} are not preserved \cite{galvao}. To construct a correct canonical quantization, we implement the gauge fixing called no-mixing $\Omega [A] = (\sqrt{1 - a^{2} \Box}) \partial_\mu A^\mu $, preserving the lagrangian order and Lorentz invariance \cite{nomixing}. Following this idea, we rewrite the Lagrangian above as

\begin{equation}\label{llkk2}
\mathcal{L}_{GQED_3} = -\frac{1}{4}F^{\mu\nu}F_{\mu\nu} - \frac{a^2}{2} \partial^\mu F_{\mu\beta} \partial_\alpha F^{\alpha\beta} - \frac{1}{2}(1-a^2 \Box)(\partial_\mu A^\mu)^2.
\end{equation}
Thus the dynamical equations are

\begin{equation}\label{elpod}
(1 - a^2 \Box ) \Box A^{\mu} = 0.
\end{equation}
We now demonstrate how the families of solutions from the $GQED_3$ differ from the Maxwell electrodynamics. One further investigation shows these equations are built up by the functions that obey distinctive differential equations. In what follows, we obtain an orthogonal decomposition where $ A^\mu $ draws independent solutions  

\begin{equation}\begin{aligned}\label{era}
(1 - a^2 \Box  ) A^\mu (x) &= A_{Max}^\mu (x), \quad \    a^2  \Box  A^\mu (x) =  A_{Pod}^\mu (x),
\end{aligned}\end{equation}
where $A^\mu_{Pod}$ and $A^\mu_{Max}$ are the Podolsky and Maxwell gauge, respectively. Therefore, the gauge field is $A^\mu = A^\mu_{Max} + A^\mu_{Pod} $ where the equations of motion are

\begin{equation}\begin{aligned}
(1 - a^2 \Box ) A_{Pod}^\mu (x) &= 0, \quad \ \ \Box A_{Max}^\mu (x) = 0.
\end{aligned}\end{equation}
To outline the quantization in momentum space, we perform the Fourier decomposition for the free electromagnetic field \eqref{elpod} into the wave solutions

\begin{equation}\begin{aligned}\label{causedd}
A_\mu (x) = \int \frac{d^2 \textbf{p}}{(2\pi)}   \sum^3_{\lambda=1}   \bigg\{ \epsilon_\mu^\lambda (p) ( a({\bf p}) e^{ipx} + a^* ({\bf p}) e^{- i p x}) + \eta_\mu (p) \bar{a}({\bf p}) e^{i \bar{p} x} + \eta_\mu^* (p) \bar{a}^* ({\bf p}) e^{- i \bar{p} x} \bigg\}.
\end{aligned}\end{equation}
We have two spectral support with disjoint domains whose the dispersion relations are $\textbf{p}^2 + \bar{p}^2_o = a^{-2}$ and ${\bf p} = i p_o $. We take now this fourier expression to derive the gauge invariant relations at equal times\footnote{ The nonvanishing relations are $[a^\lambda ({\bf p}), \ a^{* \lambda'} ({\bf p}') ] = \delta_{{\bf p},{\bf p}'} \delta^\lambda_{\lambda'} = [\bar{a}^\lambda ({\bf p}), \ \bar{a}^{* \lambda'} ({\bf p}') ]$ and the massless and massive polarizations vectors are $\epsilon_\mu^\lambda(p)$ and $\eta_\mu (p)$, respectively, with $\epsilon
^{\mu \lambda} (p) \epsilon_\mu^{\lambda'} (p) = \delta^{\lambda' \lambda}$ and $\eta^\mu (p) \eta_\mu^{*} (p) = - 1$.}     

\begin{equation}\begin{aligned}\label{qwe2}
[A_\mu (x), A_\nu (x')] &= - i \delta_{\mu\nu} D_P (x'-x), \\
\langle 0 | \{ A_\mu (x), A_\nu (x')  \} | 0 \rangle &= \delta_{\mu\nu} D^{(1)}_P (x'-x).
\end{aligned}\end{equation}
The signature of the $GQED_3$ symmetries restricts the singular functions form. By definition, we introduce the retarded $D_P^R (x) = - \Theta(x_o) D_P(x)$\footnote{We define $\epsilon (p) \equiv \frac{p_o}{|p_0|}$ and $ 2 \Theta (p) \equiv  1 + \epsilon (p) $.} and advanced  $D_P^A (x) = \Theta(- x_o) D_P (x)$ functions, the tools necessary to develop our perturbative model in the next section

\begin{subequations}\begin{align}
\label{retb}
D^R_{P} (x) &=  \int \frac{d^3 p}{(2 \pi)^3}  e^{ipx} \bigg( \mathcal{P} \frac{1}{p^2} - \mathcal{P} \frac{1}{p^2 + m_P^2} + i \pi \epsilon(p) (\delta^3 (p^2) - \delta^3 (p^2 + m_P^2))  \bigg), \\ \label{avcb}
D^A_{P} (x) &=  \int  \frac{d^3 p}{(2 \pi)^3}  e^{ipx} \bigg( \mathcal{P} \frac{1}{p^2} - \mathcal{P} \frac{1}{p^2 + m_P^2} - i \pi \epsilon(p) (\delta^3 (p^2) - \delta^3 (p^2 + m_P^2)) \bigg).
\end{align}\end{subequations}
The function $D^{(1)}_P (x)$ is 

\begin{equation}
D^{(1)}_P (x) = \int \frac{d^3 p}{(2\pi)^2} e^{ip x} (\delta^3 (p^2) - \delta^3 (p^2 + m_P^2)).
\end{equation}
What motivates a careful study is this new modes regimes $m^2 \leq p^2 \leq m_P^2$. We shall emphasize the Podolsky parameter $m_P^2$ is a natural cutoff that improves the validity of $QED_3$ to the length scale $a$. We will see this parameter plays the central role to get rid of the divergences.


\section{Heisenberg perturbative method}\label{444}

Attempts to examine quantum fluctuations lead us to several ways to describe the perturbative behavior of quantum particles. The systematic understanding of these models provides different valuable insights over quantum many-body systems. In particular, the mathematical structure involved turns them out essentially distinct even if they may arrive at the same radiative corrections. We shall remark the interpretation of the quantum field objects depends on consistent theoretical principles. In this section, we will discuss the general perturbative development in the Heisenberg picture. First, we consider the Lagrangian \eqref{llkk2} with the Dirac symmetrized and a minimal coupling sector given by

\begin{equation}\begin{aligned}\label{asdfghjk}
\mathcal{L} &=-\frac{1}{4}F^{\mu\nu}F_{\mu\nu} - \frac{a^2}{2} \partial^\mu F_{\mu\beta}\partial_\alpha F^{\alpha\beta} - \frac{1}{2}(1-a^2 \Box)(\partial_\mu A^\mu)^2 + j_\mu A^\mu \\ 
& \quad - \frac{1}{4} [( \gamma \cdot \partial + m ) \psi ,\bar{\psi}]  - \frac{1}{4} [\psi , ( - \gamma \cdot \overleftarrow{\partial} + m ) \bar{\psi} ],
\end{aligned}\end{equation}
where $j_\mu (x)$ is a matter field source. Because of the Lorentz invariance, one realizes the fermion sector and minimal coupling are the same structure of $QED_3$ \cite{Podolskyorigin}. The expected equations of motion are

\begin{subequations}
\begin{align}
(\gamma \cdot \partial + m)  \psi(x) &= i e \gamma^\mu A_\mu (x) \psi (x),  \label{cde2} \\
(1 - a^2 \Box )  \Box A^\mu (x) &=  - j^{\mu}(x). \label{cde3} 
\end{align}
\end{subequations}
It is possible to write down a current operator where $\psi$ obeys the interacting \cref{cde2,cde3} rather than the free Dirac equation. Under this argument, we symmetrize the current operator as

\begin{equation}\begin{aligned}\label{j0}
j_{\mu}(x)&=\frac{ie}{2}[\Bar{\psi}(x),\gamma_\mu \psi (x)].
\end{aligned}\end{equation}
This gauge invariant current automatically ensures $ \langle0|j^\mu(x)|0\rangle = 0 $. We can find the general solutions of inhomogeneous differential equations

\begin{equation}\label{p1}
\psi(x) = \psi^{(0)}(x) - \int d^3 x'  S_R (x-x') i e \gamma_\mu A^\mu (x') \psi(x'), \end{equation}
\begin{equation}\label{p3}
\ \ \ A_\mu (x) = A^{(0)}_\mu (x)-\int d^3 x'   D^R_P (x-x') \frac{ie}{2} [\Bar{\psi} (x'), \gamma_\mu \psi(x')], 
\end{equation}
where $(\psi^{(0)}, A^{(0)}_\mu )$ are the solutions for non-interacting \cref{cde2,cde3} and $S_R (x-x')$ follows the Schwinger's notation \cite{schwinger}\footnote{By definition, the distributions are   
\begin{equation}\begin{aligned}\label{qwe2s}
\left\{ \Bar{\psi}_a(x),  \psi_b (x') \right\} = - iS_{ba}(x'- x),\qquad \langle 0 | [\Bar{\psi}_a (x) ,  \psi_b (x')] | 0 \rangle  &= S^{(1)}_{b a} (x'- x).
\end{aligned}\end{equation}}. These results suggest the interacting operators can preserve all the original symmetries through a covariant integral formulation.

The conventional method in the Interaction picture can be briefly summarized as the direct sum of $H_0 + H_{int}$. Since the free and interacting operators lie in the orthogonal Hilbert spaces, we confront the Interaction picture rules out no unitary equivalence between the interacting and asymptotic canonical commutation relations. It follows that $H_{int}$ cannot have a well-defined ground state which obeys the vacuum symmetry of Fock space, which is quite different from the ground state of $H_0$. Thus, even though $(\psi^{(0)}, A^{(0)}_\mu)$ diagonalize $H_0$, we shall not assume $(\psi,A_\mu)$ give exactly a diagonal representation for $H_{int}$. In addition, the Euclidean transformation from $(\psi^{(0)}, A^{(0)}_\mu)$ to $(\psi, A_\mu)$ is undermined \cite{chatooo}. In this circumstance, the Interaction picture has a physical difficulty which affects the relativistic covariant result in all stages \cite{consegui}. However, the eqs. \eqref{p1} and \eqref{p3} show the interacting and free operators are defined in all Hilbert space, so the operators are unitarily equivalent to a Fock representation.


Before applying the perturbative method, we proceed with the supposition of two statements: small gauge coupling and analyticity at the origin. Following these requirements, we can develop a perturbative apparatus to analyze our local quantized model. By expanding $(\psi , A_\mu )$ into a power series of gauge coupling, we have



\begin{equation}\begin{aligned}\label{P1}
\psi(x) =  \psi^{(0)}(x) \ + \ e\psi^{(1)}(x) \ + \ e^{2} \psi^{(2)}(x) \ +\ldots, \\ 
A_{\mu}(x)  =  A_{\mu }^{(0)}(x) \ + \ e A_{\mu }^{(1)}(x) \ + \ e^{2} A_{\mu}^{(2) } (x)+\ldots,
\end{aligned}\end{equation}
Substituting \eqref{P1} into \cref{cde2,cde3}, we obtain interactively the recursive relation of the Heisenberg operators

\begin{subequations}\begin{align}\label{RRpsi}
\psi^{(n+1)}(x)
&=   - \frac{i}{2}\int{d^3 x'}  S_R(x-x')\gamma^\mu\sum_{m=0}^{n}\{A_{\mu }^{(m)}(x'),\psi^{(n-m)}(x')\}, \\ \label{RRApsi}
A_{\mu }^{(n+1)}(x)
&=  \frac{i}{2}\int{}d^3 x'  D_P^R (x-x') \sum_{m=0}^{n} [\bar{\psi}^{(m)}(x')\gamma_{\mu},\psi^{(n-m)}(x') ].
\end{align}\end{subequations}
Retaining the perturbative scheme developed so far, we may extend to the charge current 

\begin{equation}\label{EXPCurren}
j_{\mu} (x) = j^{(0)}_{\mu} (x) \ + \ e j^{(1)}_{\mu} (x) \ + \ e^2 j^{(2)}_{\mu} (x) \ + \ \ldots.
\end{equation}
Our primary concern is to retain a matrix element whose transition resumes a closed-loop correction at $e^2$ order. The lowest-order nonvanishing contribution from \eqref{j0} via \cref{RRpsi,RRApsi} is 

\begin{equation}\label{0j2}\begin{aligned}
&j^{(2)}_{\mu} (x) = \\  
& \quad \ \ \ \frac{i}{8}\int{d^3x'}\int{d^3x''} [ \bar{\psi}^{(0)}(x),\gamma_{\mu} S_R(x-x')\gamma_\nu \{ \psi^{(0)}(x') , [ \bar{\psi}^{(0)}(x''), \gamma^\nu \psi^{(0)}(x'') ] \} ]  D^R_{P}(x'-x'')\\
& \ \ - \frac{i}{4}\int{d^3x'}\int d^3x'' [ \bar{\psi}^{(0)}(x), \gamma_{\mu} S_R(x-x') \gamma_{\nu_1} S_R(x'-x'')  \gamma_{\nu_2} \psi^{(0)}(x'') ] \{ A_{\nu_1}^{(0)}(x') , A_{\nu_2 }^{(0)}(x'')   \} \\
& \ \ - \frac{i}{4} \int{d^3x'} \int d^3x'' [ \bar{\psi}^{(0)}(x') \gamma^{\nu 1} S_A(x'-x), \gamma_\mu S_R(x-x'') \gamma^{\nu 2} \psi^{(0)}(x'') ]  \{ A_{\nu_1}^{(0)}(x') , A_{\nu_2 }^{(0)}(x'')   \}\\
&  \ \  + \frac{i}{8}\int{d^3x'} \int{d^3x''}[ \{ [\bar{\psi}^{(0)}(x''), \gamma_{\nu} \psi^{(0)}(x'') ], \psi^{(0)}(x') \} \gamma^\nu  S_A(x'-x), \gamma_\mu \psi^{(0)}(x)] D^R_{P}(x'-x'') \\
& \ \ -\frac{i}{4}\int{d^3x'}\int{d^3x''} [ \bar{\psi}^{(0)}(x'')\gamma^{\nu_2}S_A(x''-x')\gamma^{\nu_1}S_A(x'-x),\gamma_{\mu}\psi^{(0)}(x) ] \{ A_{\nu_1}^{(0)}(x') , A_{\nu_2 }^{(0)}(x'') \}.
\end{aligned}\end{equation}
Although the convenient inclination deals using Feynman graphics for writing the interaction, we cannot assure one-to-one correspondence with the Feynman-Dyson $S$-matrix expansion \cite{Dyson}. We shall be aware such a statement has not been explicitly proven even though we have an agreement with few terms of the $S$-matrix series. For this reason, we appeal to build a mathematical rather than a graphic representation. Using Wick's theorem \cite{Weinberg}, we can rewrite the correction above as

\begin{equation}\label{0nnn1}\begin{aligned}
e^3 \langle q |j^{(2)}_{\mu} (x) | q' \rangle
&= \frac{ie}{2}   \int {d^3x'} \int d^3x'' \langle q | [ \bar{\psi}^{(0)}(x), \gamma_{\mu} S_R(x-x')  \sum(x'-x'') \psi^{(0)}(x'') ] |q' \rangle \ +  \\
&\quad \ \frac{ie}{2}  \int {d^3x'} \int d^3x'' \langle q | [ \bar{\psi}^{(0)}(x'') \sum(x'' -x')  S_A(x'-x), \gamma_\mu \psi^{(0)}(x) ] | q' \rangle \ +  \\
& \quad \ \frac{ie}{2} \int{d^3x'} \int d^3 x''  \langle q | [ \bar{\psi}^{(0)}(x'), \Gamma_\mu (x'-x, x - x'') \psi^{(0)}(x'')] | q' \rangle \ + \\
&\quad \ \frac{ie}{2} \int{d^3x'} \int d^3 x'' \ \Pi_{\mu\nu} (x - x')  D^R_{P}(x'-x'') \langle q |  [\bar{\psi}^{(0)}(x''), \gamma^{\nu} \psi^{(0)}(x'') ] | q' \rangle,  
\end{aligned}\end{equation}
and the same apply to $e^3 \langle 0|j^{(2)}_{\mu} (x)| q,q' \rangle$. Paying attention to each component, we can realize the first two terms correct the fermions propagator, the third one involves the vertex correction, and the last one is the vacuum polarization. We will address these points in the following sections.

\section{Vertex correction $\Lambda^\mu$ }\label{trois}

In this section, we clarify the several advantages of $GQED_3$ to examine the vertex corrections. Starting with the third component of \eqref{0nnn1} written explicitly

\begin{equation}\begin{aligned}\label{4433}
& \qquad \Gamma_\mu(x'-x, x-x'') = - \frac{e^2}{2} \gamma^\lambda Tr[  S^{(1)}(x'-x)\gamma_\mu S_{R} (x - x'') D^R_{P}(x''-x') \ + \\ \\ 
& S_A(x'-x)\gamma_\mu S^{(1)}(x-x'') D^R_{P}(x''-x') \ + \ S_A(x'-x)\gamma_\mu S_R (x-x'') D^{(1)}_{P}(x''-x')] \gamma_\lambda.
\end{aligned}\end{equation}
This distribution, invariant under translation, was first obtained in $(3+1)$ dimensions by \cite{schwinger}. However, this translational symmetry can no longer be well-defined in the Interaction picture. Let us examine the splitting operation of the total Hamiltonian $H = H_o + H_{int} $. For simplicity, we consider $\mathcal{U}$ as an unitary operator from the Euclidean group and the ground state as the unique Poincare invariant state \cite{consegui}. Since the Interaction picture assumes the dressed vacuum $| \Omega \rangle $ is proportional to the bare vacuum $| 0 \rangle$ apart from a phase factor, we can only assign a non-Fock representation to describe the interacting operator. This can be understood as follows. $| \Omega \rangle$ must coincide with the ground state of the Fock state $| 0 \rangle$, invariant under Euclidean transformation, since $ \mathcal{U} | \Omega \rangle = | \Omega \rangle $ \cite{Weinberg}. Thus the Interaction picture undermines a non-trivial system.

The Heisenberg picture presents a model based on the superposition of homogeneous and non-homogeneous solutions in \cref{p1,p3}. Then we can define a vacuum state $| \Omega \rangle$ which turns to the non-particle Fock state by an adiabatic process\footnote{We will exploit a satisfactory application in section \ref{quatre}. We shall point out the ingoing, outgoing, and free field preserve the same canonical commutation relation for finite times in the Heisenberg picture \cite{Yang}.}. To get more information from $\Gamma^{\mu}$, we introduce a Fourier transformation. 


\begin{equation}\label{cmd}
\Gamma_\mu (q,q') =  \int d^3 x' d^3 x'' e^{-iq(x' - x)}e^{-iq'(x-x'')} \ \Gamma_\mu(x'-x, x-x''). 
\end{equation}
Remembering the fermions are on-shell, we can simplify this integration\footnote{It is easy to see the $\delta^3 (k^2) \delta^3 ((q-k)^2 + m^2)$ cannot be determined simultaneously if moving on to $\vec{q}= 0$ frame.}. What remains are the terms proportional to the shift momentum in the asymptotic state. One finds

\begin{equation}\label{cmddmc}
 \Gamma_{\mu}(q,q') =  \Gamma^{(1)}_{\mu}(q,q')  +  i  \epsilon(q'-q) \Gamma^{(2)}_{\mu}(q,q').   
\end{equation}
For the sake of clarity, the physical information of these terms will be postponed. The first integral is 

\begin{equation}\begin{aligned}\label{liu}
\Gamma^{(1)}_{\mu}(q,q') &= \frac{-e^2}{8\pi^2} \int d^3 k \bigg[ \mathcal{P} \frac{\delta^3 (k^2) - \delta^3 (k^2 + m_P^2) }{((q-k)^2 + m^2)((q'-k)^2 + m^2)} + \mathcal{P} \frac{ m_P^2 \  \delta^3 ((q'-k)^2 + m^2)}{((q-k)^2 + m^2) k^2 (k^2 + m_P^2)} \\
&  \quad + \mathcal{P} \frac{  m_P^2 \ \delta^3 ((q-k)^2 + m^2) }{((q'-k)^2 + m^2) k^2 (k^2 + m_P^2) }\bigg] \gamma^\lambda(i\gamma(q-k)-m)\gamma_\mu(i\gamma(q'-k)-m)\gamma_\lambda,
\end{aligned}\end{equation}
and the second one 

\begin{equation}\begin{aligned}\label{réspondre}
\Gamma^{(2)}_{\mu}(q,q') &=  \frac{-e^2}{8 \pi} \int d^3 k \ \delta^3((q - k)^2 + m^2)\delta^3((q' -k)^2 + m^2) \bigg[  \mathcal{P} \frac{1}{k^2} - \mathcal{P} \frac{1}{k^2 + m_P^2} \bigg]\times \\
&  \ \ \ \ \ [1 - \epsilon(q' - k)\epsilon(q- k)] \gamma^\lambda (i\gamma (q-k) - m)\gamma_\mu (i\gamma (q'-k)- m )\gamma_\lambda.
\end{aligned}\end{equation}
In contrast to any fancy $QED_3$ regularization scheme, the subtraction procedure of the $GQED_3$ keeps the gauge invariance and naturally regulates the ultraviolet divergence. It is more convenient to solve \eqref{réspondre} separately in the scalar, vector, and tensorial integral. For practical reason, the tensor is

\begin{equation}\begin{aligned}\label{nettoies}
&  \mathcal{P} \int  \frac{ d^3 k \ m_P^2 \ k_\mu k_\nu }{k^2(k^2 + m_P^2)}  \delta^3((q- k)^2 + m^2)\delta^3((q' - k)^2 + m^2)[1 - \epsilon(q' - k)\epsilon(q- k)]  = \frac{ \pi \Theta (Q_o)}{\sqrt{-Q^2}} \frac{m_P^2}{Q^2}  \\ 
& \times  \frac{\Theta (- Q^2 - 4 m^2)}{Q^2 (1 + \frac{4 m^2}{Q^2})^2 } \bigg\{ \bigg[  ( q'_\mu q'_\nu + q_\mu q_\nu - \frac{Q^2}{2} g_{\mu\nu} ) -( q_\mu q'_\nu + q'_\mu q_\nu +  \frac{Q^2}{2} g_{\mu\nu} ) +  g_{\mu\nu} ( Q^2  - 4 m^2 ) \bigg]  \\ 
& \times \frac{ ( Q^2 + 4m^2 + m_P^2 ) }{\sqrt{ m_P^4 + m_P^2 Q^2 ( 1 + \frac{4 m^2}{Q^2})}} + \bigg[  ( q'_\mu q'_\nu +  q_\mu q_\nu - \frac{Q^2}{2} g_{\mu\nu} ) - ( q_\mu q'_\nu + q'_\mu q_\nu + \frac{Q^2}{2} g_{\mu\nu} )  \bigg]  \bigg\}, 
\end{aligned}\end{equation}
where $Q = q'- q$ and the vector integral is

\begin{equation}\begin{aligned}\label{eiporra3}
& \mathcal{P} \int \frac{  d^3 k \ k_\mu \ m_P^2 }{ k^2 (k^2 + m_P^2)}  \delta^3((q- k)^2 + m^2)\delta^3((q' - k)^2 + m^2)[1 - \epsilon(q' - k) \epsilon(q- k)] = \frac{\pi \Theta(Q_o)}{\sqrt{-Q^2}} \frac{m_P^2}{Q^2} \\ 
& \times  \frac{\Theta (- Q^2 - 4 m^2)}{ (1 + \frac{4m^2}{Q^2})}  
\frac{(q' + q)_\mu }{\sqrt{m_P^4 + m_P^2 Q^2 ( 1 + \frac{4 m^2}{Q^2}) }}.  
\end{aligned}\end{equation}
Before solving the scalar integral, we must regulate the infrared divergence by inserting a fictitious photon mass $(\mu)$ or \textit{t’ Hooft mass} to prevent the infrared problem in the region near $k=0$,

\begin{equation}\begin{aligned}\label{xxaswq}
& \mathcal{P} \int \frac{d^3 k \ m_P^2}{(k^2 + \mu^2)(k^2 + m_P^2) }  \delta^3((q- k)^2 + m^2)  \delta^3((q' - k)^2 + m^2)[1 - \epsilon(q' - k) \epsilon(q- k)] \\ 
& = \frac{\pi \Theta(Q_o)}{\sqrt{ -Q^2}}  \Theta (-Q^2 - 4 m^2) \bigg[ \frac{ 1}{ \sqrt{ \mu^4 + \mu^2  Q^2 ( 1 + \frac{4 m^2}{Q^2})}} - \frac{1}{ \sqrt{ m_P^4 + m_P^2 Q^2 ( 1 + \frac{4 m^2}{Q^2})}} \bigg].
\end{aligned}\end{equation}
With the aid of previous results and performing the $\gamma-$matrices algebra, we rearrange \eqref{réspondre} into two imaginary form factors $F_{1 P}$ and $F_{2 P}$

\begin{equation}\label{oreille}
\Gamma^{(2)}_{\mu}(q,q') = \gamma_\mu F_{1 P} (Q^2)  +
 i \frac{(q + q')_\mu }{2 m}  F_{2 P} (Q^2),
\end{equation}
where the parametrization suggests distinctive physical meaning, which correspond to the emission and absorption of virtual particles. The former is  

\begin{equation}\begin{aligned}\label{maxwell}
F_{1 P} (Q^2) &= -\frac{e^2}{4 \pi} \frac{\Theta(Q_o) (Q^2 + 8 m^2)}{  \sqrt{ -Q^2}} \bigg[  \frac{1}{ \sqrt{ \mu^4 + \mu^2  Q^2 ( 1 + \frac{4 m^2}{Q^2})}}-\frac{1}{ \sqrt{ m_P^4 + m_P^2 Q^2 ( 1 + \frac{4 m^2}{Q^2})} } \bigg] \\
& \quad \times  \Theta (-Q^2 - 4 m^2), \\ 
\end{aligned}\end{equation}
represents a virtual photon emitted and absorbed by a single electron. The latter is the virtual photon created and absorbed by $| q' \rangle$ and $| q \rangle$, respectively,

\begin{equation}\begin{aligned}
F_{2 P} (Q^2) &= - \frac{e^2}{3 \pi }  \frac{\Theta (Q_o)}{\sqrt{-Q^2}} \frac{m^2}{Q^2} \frac{ m_P^2}{\sqrt{m_P^4 + m_P^2 Q^2 ( 1 + \frac{4 m^2}{Q^2}) }} \frac{\Theta (- Q^2 - 4 m^2)}{(1 + \frac{4m^2}{Q^2})}.
\end{aligned}\end{equation}
Up to this point, what remains is $\Gamma^{(1)}_\mu $. We immediately infer which is not necessary to solve \eqref{liu} explicitly. We can express this integral directly similar to \eqref{oreille} without losing information about the radiative corrections. Here, as usual, the foregoing discussion used in \eqref{oreille} turns $\Gamma^{(1)}_{\mu}$ equals to

\begin{equation}\label{lambbbb}
\Gamma^{(1)}_{\mu}(q,q')  =  \gamma_\mu \bar{F}_{1 P}(Q^2) + i \frac{ (q + q')_\mu }{2m} \ \bar{F}_{2 P}(Q^2).
\end{equation}
Since $\Gamma_\mu (q,q')$ poles lie in the lower half of the complex plane and the analyticity proprieties, we can use the Kramers–Kronig relations between \eqref{oreille} and \eqref{lambbbb} \cite{Weinberg}, 

\begin{equation}\begin{aligned}
\bar{F}_{i P} (Q^2) &= \int^{\infty}_0 ds \frac{ F_{i P} (-s)}{(s + Q^2)},  
\end{aligned}\end{equation}
where $i = \{ 1,2 \} $. The form factors are 

\begin{equation}\begin{aligned}\label{lebal1}
\bar{F}_{1 P} (Q^2) &=   \frac{3 e^2}{4 \pi \sqrt{-Q^2}} \bigg[  \ln{\bigg( \frac{ 1 - \sqrt{\frac{-Q^2}{4m^2}}}{ 1 + \sqrt{\frac{-Q^2}{4m^2}}} \bigg) } - \ln{ \bigg( \frac{ 1 - \sqrt{\frac{-Q^2}{4m^2+m_P^2}}}{1 + \sqrt{\frac{-Q^2}{4m^2+m_P^2}} } \bigg) }   - \frac{8m^2}{Q^2 + 4 m^2} \sqrt{\frac{-Q^2}{4m^2}} \\
& \quad + \frac{8m^2}{Q^2 + 4 m^2 + m_P^2} \sqrt{\frac{-Q^2}{4m^2}} \bigg ] ,\\
\end{aligned}\end{equation}
and

\begin{equation}\begin{aligned}\label{lebal}
\bar{F}_{2 P} (Q^2) &=\frac{(em)^2}{3 \pi} \sqrt{\frac{m_P^2}{- Q^2}}  \bigg[ \frac{1}{(Q^2 + 4 m^2)}  \frac{\ln{\bigg( \frac{1 - \sqrt{\frac{- Q^2}{4 m^2 + m_P^2}}}{1 + \sqrt{\frac{- Q^2}{4 m^2 + m_P^2}}} \bigg)  }}{\sqrt{Q^2 + 4 m^2 + m_P^2 }} + \frac{1 - \sqrt{\frac{4 m^2}{4 m^2 + m_P^2}}}{( 3 Q^2 - 4 m^2 + m_P^2 ) m_P^2}  \\
& \quad \times \int^{\infty}_{4 m^2} \frac{ds}{\sqrt{s}}\frac{1}{(s - 4 m^2)}  \bigg],
\end{aligned}\end{equation}
where this integral shows a strong infrared divergence. The dimensional particularity in $(2+1)$ dimensions introduces only kinematics corrections for the fermion sector, absent of the anomalous magnetic momentum $\bar{F}_{2 P} (0) = 0 $.  In particular, taking the transfer momentum limit $ Q \rightarrow 0$, the vertex correction reshapes the fermion propagator by a multiplicative constant

\begin{equation}\label{lebal22}
\bar{F}_{1 P}(0)= - \frac{3}{4\pi m} \bigg(1+\frac{4 m^2}{4m^2 + m_P^2} \bigg).  
\end{equation}
We recover the $QED_3$ corrections observed in the Heisenberg picture \cite{hope} and Epstein-Glaser causal theory \cite{Thesei} by taking the limit $m_P \rightarrow \infty$. On the other hand, this operation significantly differs if we apply this limit to \eqref{lebal}. We outline it is a clear manifestation where higher derivative theory goes back to the lower one, but it requires a kinematical solution as argued in \eqref{era}. This proposition works because the planar spin has kinematical characteristics at $e^2$ order instead of dynamical one \cite{rcscsm}. The factor form gets a finite contribution due to the Podolsky term.


\section{Electron self-energy $\sum$ }\label{quatre}

As a first step towards a model where the electron interacts with its owns electromagnetic field in the Heisenberg picture, we shall describe the perturbative expansion \eqref{P1}. Following this idea, we observe how the leading ultraviolet divergences vanishes due to the Podolsky spectral support. The starting point is the matrix element, encoding all radiative corrections for the electron propagator 

\begin{equation}\begin{aligned}\label{first}
\langle 0 |\psi(x) | q \rangle &= \langle 0 |\psi^{(0)}(x) | q \rangle \ + \ e \langle 0 |\psi^{(1)}(x)| q \rangle \ + \ e^{2}\langle 0 |\psi^{(2)}(x)| q \rangle \ + \ \ldots, \\  
\end{aligned}\end{equation}
where the first element is the free propagator matrix, $\langle 0 |\psi^{(1)}(x)| q \rangle$ vanishes by Furry's theorem \cite{Weinberg}, and $\langle 0 |\psi^{(2)}(x)| q \rangle$ denotes the leading self-energy correction. Hence, using \eqref{RRpsi}, we can determine this expansion as

\begin{equation}\begin{aligned}\label{first2}
\langle 0 |\psi(x) | q \rangle &= \langle 0 |\psi^{(0)}(x) | q \rangle + \int d^3 x' S_R(x-x')\langle 0 |\Phi(x') | q \rangle + \ldots,
\end{aligned}\end{equation}
where the operator $\Phi(x')$\footnote{ We provide the explicit calculation of $\Phi(x)$ in the appendix \ref{appa}.} describes the quantum fluctuations. Before proceeding further, it is convenient to analyze this operator in a slightly different way. Assuming the weak adiabatic limit, we have

\begin{equation}\begin{aligned}\label{hhh}
& \langle 0 |\Phi (x,\alpha) | q \rangle =\\ 
& \qquad \ i \int \frac{d^3p}{(2\pi)^2}   \int^x_{-\infty} d^3x' e^{\alpha(x_o + x_o') + ip(x-x')} \epsilon(p) \bigg[ \sum_1(p^2) + (i\gamma p + m) \sum_2(p^2) \bigg]  u(q) e^{i q x'}, 
\end{aligned}\end{equation}
where the function $e^{\alpha(x_o + x_o')}$ smooths out the integral and reproduces the adiabatic hypothesis by the parameter $\alpha$ 

\begin{equation}\begin{aligned}\label{fffdd}
& \langle 0 |\Phi (x,\alpha) | q \rangle  = \\
& \qquad i e^{2 \alpha x_o} \int \frac{dp_o}{\alpha + i(p_o-q_o)} \epsilon(p) \bigg[ \sum_1(\textbf{q}^2 - p^2_o) + (i\gamma_k p_k - \gamma_4 p_o  + m) \sum_2(\textbf{q}^2 - p^2_o) \bigg] u(q) e^{iqx}.
\end{aligned}\end{equation}
It is important to note that the energy distribution of \eqref{fffdd} reflects the cloud of virtual photons surrounding the electron. A suitable transformation of variables $\textbf{q}^2 - p^2_o = -s$ gives

\begin{equation}\begin{aligned}\label{435534}
\langle 0 |\Phi (x,\alpha) | q \rangle &=  e^{2 \alpha x_o } \int^{\infty}_0   \frac{ds}{s + \textbf{q}^2 -(q_o + i\alpha)^2} \bigg[ \sum_1(-s) - i \alpha \gamma_4 \sum_2(-s) \bigg] \langle 0 | \psi^{(0)}(x) | q \rangle.
\end{aligned}\end{equation} 
Taking the adiabatic limit $\alpha \rightarrow 0$ turns this matrix element into the asymptotic state

\begin{eqnarray}\label{cacilda} 
\langle 0 |\Phi (x) | q \rangle =  \bar{\sum_1} (-m^2) \langle 0 |\psi^{(0)}(x)| q \rangle.
\end{eqnarray}
This process produces the transition from the Hamiltonian $H ( \psi, A_\mu )$ to the asymptotic one $H ( \psi^{out} , A_\mu^{out} ) $\footnote{We can construct the ingoing and outgoing Hamiltonian by the free one at $x_o \rightarrow + \infty$ and $x_o \rightarrow - \infty$, respectively. They satisfy the same commutation relations and differential equations of motion \cite{Yang}.} where $\alpha$ allows us to distinguish what region of the radiative process is taken into account. The requirement of such transition gives a well-defined self-adjoint operator on the Fock space, compatible with a unitary representation of the Poincaré group \cite{acabalogo}. Understanding this point is essential to deal with the electron propagator correction. 
It is also interest to analyze \eqref{first2} by multiplying the Dirac equation on both sides

\begin{equation}\label{casado}
(\gamma \cdot \partial + m )  \langle 0 |\psi(x) | q \rangle = -  \langle 0 | \Phi (x) | q \rangle.
\end{equation}
If we set up $q^2 = -m^2$, we obtain the asymptotic state



\begin{equation}\begin{aligned}\label{cdzqed}
& (\gamma \cdot \partial + m )  \langle 0 |\psi (x) | q \rangle \\
& = - \bigg[ \bar{\sum_1}(-m^2)  +  i \epsilon(q) \sum_1(-m^2)  +  (i\gamma q + m) \bigg(  \bar{\sum_2}(-m^2)  + i \epsilon(q)\sum_2(-m^2) \bigg) \bigg] 
\langle 0 |\psi^{(0)}(x) | q \rangle, \\
&= - \bar{\sum_1}(-m^2)   \langle 0 |\psi^{(0)}(x)| q \rangle.
\end{aligned}\end{equation}
It is not surprising we find this result in complete agreement with  \eqref{435534}. Our method intends to obtain a finite equation of motion for a single electron. To interpret such an assumption, we shall make a careful study of the divergent quantities. Thus we need to figure out what parameters are necessary to get finite values in \eqref{cdzqed}. Let us see how to get over this difficulty. First, it is wise to rewrite this equation to give a complete equivalent "free" Dirac equation at $e^2$ order

\begin{equation}\label{tre}
(\gamma \cdot \partial + m + \bar{\sum_1}(-m^2) ) \langle 0 |\psi (x) | q \rangle = 0,
\end{equation}
where the bare mass $m$ is an infinity parameter at $x \rightarrow - \infty$. It becomes clear we shall interpret the self-energy function $\bar{\sum_1}(-m^2)$ as the observed mass minus bare one. This operation, called on shell renormalization scheme, enables us to rewrite \eqref{tre} as

\begin{equation}\label{tre3}
(\gamma \cdot \partial + m_{pole})\langle 0 |\psi^{(r)} (x)|q \rangle=0,
\end{equation}
where $\psi^{(r)}$ is a finite renormalized operator and $m_{pole}$ is the finite pole or observed mass. Then we write this differential equation of motion with finite quantities, encoding the leading radiative corrections\footnote{To understand what is happening, this treatment yields the same renormalized Green's function $( \gamma \cdot \partial + m_{pole}) S_R (x)= -\delta^3 (x)$, which display a propagator with pole $ \slashed{p} = i m_{pole}$ and residue $-1$.}. What remains to discuss is the relation between the renormalized and bare operator

\begin{equation}\label{5577}
\langle 0|\psi^{(r)} (x)|q \rangle = Z_2^{-1/2} \langle 0|\psi^{(0)} (x)|q\rangle,
\end{equation}
where $Z_2 $ is a finite constant. As we have seen in \eqref{cacilda}, $\Phi(x)$ is composed of the free field operators. We may tempt to apply in a reasonable way \eqref{first2} to compute $Z_2$. But such argument is misleading. One quite general grounds, we can correctly achieve the radiative correction by subtracting  $\langle0|\Phi(x,\alpha)|q\rangle$ from the asymptotic state \eqref{cacilda}


\begin{eqnarray}\begin{aligned}\label{label}
&\langle0|\Phi(x,\alpha)|q\rangle - e^{2\alpha x_o}\bar{\sum_1}(-m^2) \langle 0 |\psi^{(0)}(x)| q \rangle  =     \\
& \ e^{2 \alpha x_o } \int ds \bigg[  \frac{\sum_1(-s)}{(s-m^2)} \frac{  (2i q_o \alpha -\alpha^2)  }{(s + \vec{q}^2 -(q_o + i\alpha)^2)}
-\frac{ (i \gamma_4 \alpha)  \sum_2 (-s)}{s + \vec{q}^2 -(q_o + i\alpha)^2} \bigg]  \langle 0 |\psi^{(0)} (x)| q \rangle.
\end{aligned}\end{eqnarray}
Here, $e^{2 \alpha x_o }$ ensures the adiabatic interaction since $ \bar{\sum_1}  \rightarrow  0$ for $ e \rightarrow 0$. We now replace the expression above into \eqref{first}

\begin{equation}\begin{aligned}\label{butterf}
\langle 0 |\psi^{(r)} (x)  | q \rangle &= \bigg\{1 + i \int \frac{d^3 p}{(2 \pi)^2} \int^{x}_{-\infty} d^3 x' e^{ip(x-x')} \delta^3 (p^2 + m^2) \epsilon(p) (i \gamma \cdot p - m) e^{2 \alpha x_o } \int^{\infty}_0 ds 
 \\ 
& \quad \times  \bigg[ \frac{\sum_1(-s)}{(s-m^2)}  \frac{(2i q_o \alpha -\alpha^2)}{(s + \textbf{q}^2 -(q_o + i\alpha)^2)}  - \frac{(i\gamma_4 \alpha) \sum_2(-s)}{s + \textbf{q}^2 -(q_o + i\alpha)^2} \bigg] \bigg\} \langle 0 |\psi^{(0)}(x)  | q \rangle. \\
\end{aligned}\end{equation}
After integrating and taking the limit $\alpha \rightarrow 0$, we have 


\begin{equation}\begin{aligned}\label{ffaass}
\langle 0 |\psi^{(r)} (x)  | q \rangle 
&=  \bigg\{ 1 - \frac{1}{2} \int^{\infty}_0 ds \bigg[ \frac{\sum_2(-s)}{(s-m^2)}-2m\frac{\sum_1(-s)}{(s-m^2)^2} \bigg] \bigg\} \langle 0 |\psi^{(0)}(x)  | q \rangle.
\end{aligned}\end{equation}
Finally, we obtain the finite constant

\begin{equation}\begin{aligned}\label{zureteba}
Z^{-1/2}_2 &= 1 - \frac{1}{2} \bigg[ \bar{\sum_2}(q^2)\bigg|_{q^2 \rightarrow - m^2} + 2 m \frac{ \partial \bar{\sum_1} (q^2)}{\partial q^2}\bigg|_{q^2 \rightarrow - m^2} \bigg], \\ 
\end{aligned}\end{equation}
where we define the counterterm $(Z_2 - 1 ) \equiv   \delta_2$ up to $e^2 $ order as

\begin{equation}\begin{aligned}\label{devenir}
\delta_2 &=  - \frac{3 e^2}{8 \pi m } \bigg[ \ \bigg( 3 - \frac{m_P^2}{m^2} \bigg) \ln{\bigg( 1 + \sqrt{\frac{ 4 m^2}{m_P^2}} \bigg)} + 3 \lim_{q^2 \to -m^2} \ln{\bigg(\frac{1- \sqrt{\frac{-q^2}{m^2}}}{1+ \sqrt{\frac{-q^2}{m^2}}} \bigg)} + \sqrt{\frac{4 m_P^2}{m^2}} \bigg].
\end{aligned}\end{equation}
Even if the function is well-behaved in the ultraviolet regions, the fermion propagator develops the mass-shell singularity at the pole $\slashed{p} = i m_{pole}$ which does not guarantee the Taylor expansion of \eqref{zureteba}. Near the pole mass, the singularity structure behaves like a logarithmic divergence in \eqref{devenir} by analyzing $\delta_2$. This divergence is not moderated by $m_P$ and invalidates the gauge independence of the fermion propagator. In addition, notice this situation comes out only due to the Dirac algebra in \eqref{ffaass} which conducts an unbounded spectrum in $p^2 \rightarrow - m^2$ for $m \neq 0$. It is clear the functions $\bar{\sum_1}$ and $\bar{\sum_2}$ only exhibit the mass shell singularity for $d \leq 2+1$ dimensions. Even though the Podolsky contribution circumvents the infrared catastrophe with finite terms at $\delta_2$, the $GQED_3$ still poses an infrared problem. In other words, the geometric deformation considered in \eqref{causedd} contributes to erasing ultraviolet divergence, insensitive to the long distance of the photon propagator. As we known, the confinement of the fermion because of the strong infrared behavior cannot be possible since \eqref{devenir} undermines this asymptotic state \cite{confinement}. Although it is rather intuitive, the infrared divergence vanishes if summing all contribution in the quenched rainbow approximation \cite{Maskawa}. However, the mass shell singularity continues a problem \cite{raiva}. Returning to \eqref{cde2}, we can formalize the successive operations considered so far. By adding $ \delta m$ on both sides, we consider the \textit{mass renormalization} effect in the Heisenberg picture

\begin{equation}\label{0zzzz}
(\gamma \cdot \partial + \underbrace{ m + \delta m }_{m_{pole}} ) \psi(x)  =  i e \gamma^\mu A_\mu (x) \psi(x) + \delta m  \psi(x),
\end{equation}
where we interpret the term $\delta m  \psi(x)$ as the mass radiation because of \eqref{cdzqed}. Adding $\delta m$ in both sides may be completely redundant. We shall note this term plays a different physical role depending on which side we are dealing. The left side is independent of adiabatic process, while the right one shows $\delta m \rightarrow 0$ for $ e \rightarrow 0$. Using \eqref{tre}, the counterterm is equal to


\begin{equation}\label{ffpp}
 \delta m = - \bar{\sum_1}(-m^2).
\end{equation}
After all this ponderous work, we go back to \eqref{0nnn1}. Beginning from $ \sum (q)$, we obtained without difficulty the first two terms of \eqref{0nnn1} from \eqref{label} to \eqref{zureteba}. To calculate the third term, we use the Fourier transform in \eqref{cmddmc} with the aid of \eqref{oreille} and \eqref{lambbbb}. Finally, the last term is the vacuum polarization\footnote{
The $\gamma$-matrices in $(2+1)$ dimensions introduce a total antisymmetric component $\epsilon^{\mu \nu \alpha} $ into the general structure of polarization tensor 
\begin{equation}\label{exit}
\Pi^{\mu\nu}(Q^2) =  (g^{\mu\nu} - Q^\mu Q^\nu /Q^2 ) \Pi^{(1)}(Q^2) + i m \epsilon^{\mu \nu \alpha} k_\alpha \Pi^{(2)}(Q^2).
\end{equation}
For a comprehensive discussion at one-loop order in the Heisenberg picture, see Ref. \cite{Tomazelli4}. In contrast with $\Gamma^\mu$ and $\sum$ corrections, polarization in the $GQED_3$ gives the same result for the $QED_3$ \cite{Podolskyorigin}. As we have seen in \eqref{asdfghjk}, the photon interaction with virtual one-loop fermion is unaffected.}. In what follows, the matrix element \eqref{0nnn1}  reads

\begin{equation}\begin{aligned}\label{chegaa}
& \langle q | j^{(2)}_\mu (x) | q' \rangle   =  \langle q | j^{(0)}_\mu (x) | q' \rangle  \bigg[ -  \Bar{\Pi} (Q^2)  +  \Bar{\Pi} (0)  -  i  \epsilon (Q) \Pi (Q^2)  -   \bar{\sum_2} (-m^2) - 2 m \bar{\sum_1} (-m^2) + \\ 
&    \Bar{F}_{1 P} (Q^2)  +  i  \epsilon(Q) F_{1 P} (Q^2) \bigg]  - e \frac{(q+q')_\mu}{2m}  \bigg[  \Bar{F}_{2 P} (Q^2) + i \epsilon(Q) F_{2 P} (Q^2) \bigg] \langle  q | : \psi^{(0)} (x) \psi^{(0)} (x) :  | q' \rangle. 
\end{aligned}\end{equation}
Substituting the previous results into \eqref{chegaa} yield

\begin{equation*}\begin{aligned} 
& \langle q | j^{(2)}_\mu (x) | q' \rangle  = e^2 \frac{\langle q|j^{(0)}_\mu (x)| q' \rangle}{\sqrt{- Q^2}}  \bigg\{  \frac{1}{2 \pi } \bigg[ \bigg( 1 - \frac{4m^2}{Q^2} \bigg)  \ln \bigg( \frac{1- \sqrt{\frac{-Q^2}{4m^2}}}{1+\sqrt{\frac{-Q^2}{4m^2}}}\bigg) -  4 \sqrt{\frac{4 m^2}{- Q^2}} - \frac{i}{2} \epsilon(Q) \Theta(Q_o)\\
& \times \Theta (Q^2 + 4 m^2) \bigg( 1 - \frac{4 m^2}{ Q^2 } \bigg) - \sqrt{\frac{- Q^2}{ 4 m}} 
- \bigg[\ln\bigg(\frac{1-\sqrt{\frac{-Q^2}{4m^2}}}{1+\sqrt{\frac{-Q^2}{4m^2}}}\bigg) + \frac{i}{2} \epsilon(Q) \Theta(Q_o) \Theta(Q^2 + 4 m^2) \bigg] + \\ 
& \ \ \sqrt{\frac{-Q^2}{4 m }} \bigg[\bigg( 3 - \frac{m_P^2}{m^2} \bigg) \ln{\bigg( \frac{m_P}{2 m + m_P} \bigg)} + 3 \lim_{q^2 \to -m^2} \ln{\bigg(\frac{1- \sqrt{\frac{-q^2}{m^2}}}{1+ \sqrt{\frac{-q^2}{m^2}}} \bigg)} + \sqrt{\frac{4 m_P^2}{m^2}} \bigg]   - 3 \bigg[ \sqrt{\frac{-Q^2}{4 m^2}}   \\ 
& \times  \frac{ 8 m^2 m_P^2}{(Q^2 + 4 m^2)(Q^2 + 4 m^2 + m_P^2)}  +  \ln{ \bigg( \frac{ 1 - \sqrt{\frac{-Q^2}{4m^2+m_P^2}}}{1 + \sqrt{\frac{-Q^2}{4m^2+m_P^2}} } \bigg) } - \ln{\bigg( \frac{ 1 - \sqrt{\frac{-Q^2}{4m^2}}}{ 1 + \sqrt{\frac{-Q^2}{4m^2}}} \bigg) } \bigg] \bigg] +   i \epsilon(Q) \Theta(Q_o)  \\
& \times  \Theta (-Q^2 - 4 m^2)  \bigg[ \frac{1}{ \sqrt{ \mu^4 + \mu^2  Q^2 ( 1 + \frac{4 m^2}{Q^2})}} -   \frac{1}{ \sqrt{ m_P^4 + m_P^2 Q^2 ( 1 + \frac{4 m^2}{Q^2})} } \bigg]  (Q^2 + 8 m^2)  \bigg\}  - 
\end{aligned}\end{equation*}

\begin{equation}\begin{aligned}\label{préférer}
& \ e^3 \frac{(q+q')_\mu}{2m} \bigg\{ \frac{m^2 }{3 \pi} \sqrt{\frac{m_P^2}{- Q^2}} \bigg[ \frac{1}{(Q^2 + 4 m^2)} \frac{\ln{\bigg( \frac{1 - \sqrt{\frac{- Q^2}{4 m^2 + m_P^2}}}{1 + \sqrt{\frac{- Q^2}{4 m^2 + m_P^2}}} \bigg)  }}{\sqrt{Q^2 + 4 m^2 + m_P^2 }} +    \frac{1 - \sqrt{\frac{4 m^2}{4 m^2 + m_P^2}}}{( 3 Q^2 - 4 m^2 - m_P^2) m_P^2 } \\
&   \times \int^{\infty}_{4 m^2} \frac{ds}{\sqrt{s}}\frac{1}{(s - 4 m^2)}  \bigg] +  i \frac{ \epsilon(Q) \Theta (Q_o)}{\sqrt{-Q^2}}\frac{m^2}{Q^2}  \frac{\Theta (- Q^2 - 4 m^2)}{(1 + \frac{4m^2}{Q^2})} \frac{  m_P^2}{\sqrt{m_P^4 + m_P^2 Q^2 ( 1 + \frac{4 m^2}{Q^2})}} \bigg\}   \\
& \times \langle  q | : \psi^{(0)} (x) \psi^{(0)} (x) : | q' \rangle .
\end{aligned}\end{equation}
As we clearly showed, the $\bar{F}_{2 P} (0)$ vanishes even if we enlarge the parameter space with the Podolsky constant. A useful observation is the higher derivative framework increases the photon kinematic nature but not the spin dynamical. Then there is no shift in the electron magnetic moment at $e^2$ order. The emergence of this current depends on the fictitious mass $\mu$, which also occurs in $QED_3$ \cite{hope}. We expect this feature since the higher derivative theory does not influence the behavior of correlation function in the long-range distance. Although we can evaluate $j_\mu$ for all orders without worrying about ultraviolet divergences, the infrared region and mass shell singularity demand various peculiarities which are beyond the perturbative method discussed in ths article.   


\section{Conclusions and Perspectives}\label{doendo}

We introduced the subject of planar electrodynamics from the viewpoint of the higher derivatives theory. We discussed how the appropriate gauge choice was fundamental to drop out the unnecessary degrees of freedom from the new families of solutions \eqref{elpod}. In the formulation adopted by us, the $GQED_3$ presents a more fundamental solution than the $QED_3$, which explores the energy spectrum in the limits $m^2 \leq p^2 \leq m_P^2$. What these calculations showed is the former theory could remove the natural ultraviolet divergences from the latter one. Proceeding with this idea, we calculated the finite quantum excitations: the vertex function $\Gamma^\mu$ and electron self-energy $\sum$ by subtracting two orthogonal solutions \eqref{era}. For such a task, the Podolsky parameter, presented in this article, emerged as a natural regulator in the radiative process \eqref{préférer}. We noted our construction keeps the super-renormalizable characteristics since \eqref{asdfghjk} modifies only the kinetic property, so the minimal coupling continues described by a marginal coupling. We cleared away the first barrier of ultraviolet divergence, whereas removing the $IR$ open several questions since infrared problems are more severe in $(2+1)$ than $(3+1)$ dimensions \cite{camadade}. We showed the planar fermions exhibited in \eqref{zureteba} a mass shell singularity at one-loop, invalidating the Taylor expansion. Our theory gives a new perspective to study the non-perturbative \cite{Hoshino} and higher-order perturbative \cite{Mitra} effects on the confinement. Beyond that, there is a natural way to remove the mass shell singularity by redefining the $S$-matrix series and Fock space \cite{Tomazelli1,Tomazelli8}. We leave these methods to the future.


The theoretical analysis showed the $QFT$ principle could stand around ill-defined conceptions since the Interaction picture give us inconsistent physical interpretation. We explored the Heisenberg picture to interpret the one-loop radiative corrections. This picture is not only to circumvent the Interaction picture as a simple mathematical alternative but rather a consistent apparatus to define the quantum field objects. The main Heisenberg picture characteristic was to having operators defined within all Hilbert space, having unitary equivalence of the Fock representation and interacting fields, and a linear map between the asymptotic and free Hilbert spaces \cite{Yang}.

The antisymmetric coupling called Chern-Simons $CS$ is added explicitly in the Lagrangian to present a massive gauge in the Maxwell-Chern-Simons model $(MCS)$. As demonstrated in the Interaction picture \cite{chatoo,perturbcs}, we can adopt the $CS$ as a physical manifestation of vacuum polarization at one-loop order. We tried to argue this fundamental origin comes from the antisymmetric term in \eqref{exit}, which \textit{generates dynamically} a physical mass for the photon. Such a process is already confirmed perturbatively \cite{Tomazelli4} and non-perturbatively \cite{rabanal} in the Heisenberg picture. We are able to show this effect in \eqref{préférer} from $\Pi$. We can support this argument since the Hilbert space of the $GQED_3$ is equivalent to $MCS$ for asymptotic photon propagator \cite{rabanal}.       


We showed the planar spin has kinematical information at $e^2$ order \eqref{lebal} \cite{rcscsm}, whereas a detailed investigation at $j_\mu^{(4)}$ could arise a dynamical behavior. Our perturbative framework implements a way to extract the real part of $\bar{F}_{2 P} (0)$, responsible for the anomalous magnetic moment. Performing a correct treatment for vacuum polarization, we can substitute into \eqref{4433} the virtual photon propagator, roughly speaking, $1/k^2$ by the massive one $ 1/(k^2 + \Pi(0))$ from \eqref{exit}. This antisymmetric contribution guide us to a $P$-odd term in $\Gamma^{\mu} = \gamma^\mu + i \epsilon^{\mu \nu \sigma} \gamma_\nu q_\sigma$, which would induce a magnetic moment. This contribution affects the dynamic of anyons as argued in \cite{rcscsm} and its investigation will be published elsewhere.

We end by wondering what happens with $GQED_3$. It is interesting to list some directions which we left to apply our research. One challenge is to discover the critical temperature for the thermal fluctuation that erases the vacuum polarization. It might be useful to analyze the second-order phase transition from the chiral symmetry breaking \cite{quist}. We also hope to explore the rich structure of the condensed matter in light of $GQED_3$. This framework may incorporate accurate spectral support to yield a systematic explanation of electron-electron bound state on high$-T_c$ superconductivity \cite{mcdmcd}. Finally, we stress that much of our analysis in the Heisenberg picture is valid to explore the infrared properties by a non-perturbative aspect \cite{deRoo} and the endemic infrared divergence in the Schwinger-Dyson equation \cite{finitetem}. This analysis could shed light on the inconsistency problem of the vacuum expectation value of the condensates. These issues will be reported elsewhere.

\section*{Acknowledgements}
D. Montenegro thanks to CAPES for full support. B. M. Pimentel thanks to CNPq for partial support.

\appendix
\section*{Appendix}

\section{Electron transition amplitude}\label{appa}

Here, we outline the details behind the derivation of \eqref{hhh}. The structure of the integral follows
 
\begin{equation}\label{zorra}
\Phi(x) = - \frac{e^2}{2} \int d^3x' \gamma_\lambda [S^{(1)}(x-x') D^A_{P} (x'-x) + S_{R} (x-x') D^{(1)}_{P} (x'-x)  ] \gamma^\lambda \psi^{(0)}(x').
\end{equation}
We can propose a method, without loss of generality, to describe the operator as

\begin{equation}\label{0234}
 \Phi(x) = \int d^3x' \ \sum(x-x') \psi^{(0)}(x'),   
\end{equation}
where the $\psi^{(0)}(x')$ denotes the external source and the kernel $\sum(x-x')$ is given by \eqref{zorra}. As the function is well behaved, we shall take the Fourier transform 
  
\begin{equation}\label{realone}
\sum (x-x') = \int \frac{d^3 q}{(2 \pi)^3} \ e^{iq(x-x')} \sum (q).  
\end{equation}
To facilitate the calculation, we shall decompose the expression into the real and imaginary parts. Starting from the imaginary one

\begin{equation}\begin{aligned}\label{demandé}
Im \sum (q)  &=  \frac{e^2}{2} \int \frac{d^3 p}{(2\pi)^2}  d^3 k \ \delta^3( q - p + k)  \delta^3(  p^2  + m^2)   (\delta^3 ( k^2) - \delta^3 ( k^2 + m_P^2)) ( \epsilon(  k  ) - \epsilon( p ) )  \\
& \quad  \times \gamma^\lambda  ( i\gamma  p  - m) \gamma_\lambda.
\end{aligned}\end{equation}
In accord with the phenomenological view, we shall interpret alternatively this result as       

\begin{equation}
Im \sum (q) = \epsilon (q) \bigg[ \sum_1 (q^2)  + (i \gamma q + m) \sum_2 (q^2) \bigg]. 
\end{equation}
The term $ \epsilon(q) \sum_2 (q^2)$ can be isolated by multiplying on both sides of the gamma matrix and summing over spinorial indices. Thus, we have 
      
\begin{equation}\begin{aligned}\label{lam}
\epsilon (q)  \sum_2 (q^2) &=  \frac{e^2}{4} \int \frac{d^3 k}{(2 \pi)^2} \bigg( 1 - \frac{m^2}{q^2} \bigg) \delta^3 (q^2 +2qk + k^2 + m^2) ( \delta^3(k^2) - \delta^3(k^2 + m_P^2) )  \\
& \quad  \times (\epsilon(q+k) - \epsilon(k)).
\end{aligned}\end{equation}
Proceeding in a manner similar, we obtain the $\epsilon (q) \sum_1 (q^2)$  by summing over the spinorial indices

\begin{equation}\begin{aligned}\label{lam2}
\epsilon (q) \bigg[ \sum_1 (q^2) + m \sum_2 (q^2) \bigg] &= \frac{m e^2}{4 \pi^2} \int d^3 k  \delta^3((q+k)^2 + m^2) ( \delta^3(k^2) - \delta^3(k^2 + m_P^2) )  \\
& \quad \times (\epsilon(q+k) - \epsilon(k)).
\end{aligned}\end{equation}
By observing that $\sum (q)$ is analytic, invariant under translational symmetry, and vanishes if $q$ lies in the forward light cone. We can find the real part of \eqref{realone} through the following Kramers–Kronig relations \cite{Weinberg}

\begin{equation}\label{cackoo}
\bar{\sum_{i}}(q^2) = P \int^{\infty}_{0} ds \frac{\sum_{i} (-s)}{s+q^2},      
\end{equation}
where $i = \{ 1,2 \} $. The general expression for the electron self-energy is 

\begin{equation}\begin{aligned} 
\sum (q) &= \bar{\sum_{1}}(q^2) + i  \epsilon(q) \sum_{1}(q^2) + (i \gamma q + m)[  \bar{\sum_{2}}(q^2) + i \epsilon(q) \sum_{2}(q^2)]. \\ 
\end{aligned}\end{equation}
With the aid of \eqref{lam}, \eqref{lam2}, and \eqref{cackoo}, the self-energy at one-loop order is


\begin{equation}\begin{aligned} 
\sum (q) &= \frac{m e^2}{8 \pi \sqrt{-q^2}} \bigg\{ \bigg[ \bigg( 3 + \frac{m^2 - m_P^2}{q^2} \bigg)  \ln{\bigg( \frac{ 1 - \sqrt{\frac{-q^2}{(m+m_P)^2}}}{  1 + \sqrt{\frac{-q^2}{(m+m_P)^2}}  } \bigg) } - \bigg( 3 + \frac{m^2}{q^2} \bigg) \ln{\bigg( \frac{ 1 - \sqrt{\frac{-q^2}{m^2}}}{  1 + \sqrt{\frac{-q^2}{m^2}} }\bigg) } \\ 
& \quad -  \sqrt{\frac{ 4 m_P^2}{-q^2}}  \bigg ]  \ + \ i \pi    \epsilon(q)\frac{\Theta(q_o)}{2} \bigg[ \bigg( 3 + \frac{m^2}{q^2} \bigg) \Theta (- q^2 - m^2)  -    \bigg( 3 + \frac{m^2 - m_P^2}{q^2} \bigg)  \\
& \ \times \Theta(-q^2 - (m + m_P)^2) \bigg] \bigg\} + e^2  \frac{ (i \gamma q + m)}{8 \pi \sqrt{-q^2}}   \bigg\{ \bigg[  \bigg( 1-\frac{m^2 - m_P^2}{q^2} \bigg) \ln{ \bigg( \frac{  1 - \sqrt{\frac{-q^2}{(m+m_P)^2}}  }{  1 + \sqrt{\frac{-q^2}{(m+m_P)^2}}}\bigg)} \\ 
& \  - \bigg( 1 - \frac{m^2}{q^2} \bigg) \ln{\bigg( \frac{   1 - \sqrt{\frac{-q^2}{m^2}} }{  1 + \sqrt{\frac{-q^2}{m^2}}}\bigg)} + \sqrt{\frac{4 m_P^2 }{-q^2}} \bigg]   + i \pi \epsilon(q) \frac{\Theta(q_o)}{2} \bigg[ \bigg(1 - \frac{m^2}{q^2} \bigg)  \Theta (- q^2 - m^2) \\ 
& \ -  \bigg( 1 - \frac{m^2 - m_P^2}{q^2} \bigg)    \Theta(-q^2 - (m + m_P)^2) \bigg] \bigg\}.
\end{aligned}\end{equation}


\end{document}